# Understanding and calibrating Density-Functional-Theory calculations describing the energy and spectroscopy of defect sites in hexagonal boron nitride.

Jeffrey R. Reimers,[1,2*] A. Sajid,[2,3] Rika Kobayashi,[4] and Michael J. Ford[2*]

[1]*International Centre for Quantum and Molecular Structures and Department of Physics, Shanghai University, Shanghai 200444, China*

[2]*University of Technology Sydney, School of Mathematical and Physical Sciences, Ultimo, New South Wales 2007, Australia*

[3]*Department of Physics, GC University Faisalabad, Allama Iqbal Road, 38000 Faisalabad, Pakistan.*

[4]*National Computational Infrastructure, The Australian National University, Canberra, ACT 2600, Australia.*

**ABSTRACT:** Defect states in 2D materials present many possible uses but both experimental and computational characterization of their spectroscopic properties is difficult. We provide and compare results from 13 DFT and *ab initio* computational methods for up to 25 excited states of a paradigm system, the $V_NC_B$ defect in hexagonal boron nitride (h-BN). Studied include include: (i) potentially catastrophic effects for computational methods arising from the multi-reference nature of the closed-shell and open-shell states of the defect, which intrinsically involves broken chemical bonds, (ii) differing results from DFT and time-dependent DFT (TDDFT) calculations, (iii) comparison of cluster models to periodic-slab models of the defect, (iv) the starkly differing effects of nuclear relaxation on the various electronic states as broken bonds try to heal that control the widths of photoabsorption and photoemission spectra, (v) the effect of zero-point energy and entropy on free-energy differences, (vi) defect-localized and conduction/valence band transition natures, and (vii) strategies needed to ensure that the lowest-energy state of a defect can be computationally identified. Averaged state-energy differences of 0.3 eV are found between CCSD(T) and MRCI energies, with thermal effects on free energies sometimes also being of this order. However, DFT-based methods can perform very poorly. Simple generalized-gradient functionals like PBE fail at the most basic level and should never be applied to defect states. Hybrid functionals like HSE06 work very well for excitations within the triplet manifold of the defect, with an accuracy equivalent to or perhaps exceeding the accuracy of the *ab initio* methods used. However, HSE06 underestimates triplet state energies by on average 0.7 eV compared to closed-shell singlet states, while open-shell singlet states are predicted to be too low in energy by 1.0 eV. This leads to miss-assignment of the ground state of the $V_NC_B$ defect. Long-range corrected functionals like CAM-B3LYP are shown to work much better and to represent the current entry level for DFT calculations on defects. As significant differences between cluster and periodic-slab models are also found, the widespread implementation of such functionals in periodic codes is in urgent need.

## 1. INTRODUCTION

There is currently great interest in the spectroscopy of point defects in 2-D and 3-D semiconducting materials as this can be exploited to fabricate qubits for quantum computing technology[1-4] as well as for single photon sources for quantum cryptography. While much work has focused on the negatively charged nitrogen vacancy center ($N_V^{-1}$) in diamond,[5-6] hexagonal boron nitride (h-BN) also has the potential to host many such color centres.[7-14] Defect occurrences are not only naturally occurring[15-16] but can also be artificially induced and controlled,[17] leading to device possibilities.

The precise nature of observed defects is often difficult to determine, with potentially useful defects being difficult to predict, making popular their investigation by first principles computational techniques.[18-24] However, methods for performing accurate calculations of this type are still being developed. Small molecules containing the types of functionalities found in diamond and h-BN defects may be described to chemical accuracy using standard *ab initio* computational methods such as quantum Monte Carlo (QMC),[25] coupled-cluster singles and doubles (CCSD)[26] theory with perturbation corrections for triples excitations (CCSD(T)),[27] related time-dependent approaches such as equations of motion coupled cluster (EOMCCSD),[28-29] singles and doubles multi-reference configuration interaction (MRCI),[30] as well as its approximation, complete-active-space self-consistent field (CASSCF)[31] theory with perturbative corrections for singles and doubles excitations (CASPT2).[32] While CCSD(T) is regarded as being the "gold standard" in molecular electronic structure computation, the multi-reference nature of many of the defect states of interest could demand QMC or MRCI approaches instead. However, CCSD and MRCI based methods can only be applied to model compounds,[33] not materials such as 3-D h-BN or 2-D h-BN nanoflakes. A range of improved high-level methods is available[34-53] but unfortunately rarely deployed.

Nevertheless, the most commonly applied approaches to electronic and nuclear structure simulation in modern times include density-functional theory (DFT) and its time-dependent variant, time-dependent DFT (TDDFT). Despite having no method available for the systematic improvement of DFT calculations towards the exact answer, in practical calculations on sizeable molecules, DFT and TDDFT approaches can often deliver similar accuracy to their much more computationally expensive *ab initio* counterparts. DFT can be regularly applied to periodic systems, allowing an alternate approach to modelling defect sites. In molecular-cluster-based approaches, the size of the cluster must be increased until all long-range effects of interest are included, a difficult task if transitions involve the valence and/or conduction bands of the solid, whereas for periodic-solid approaches, the size of the unit cell must be increased until neighboring defect sites do not interfere with each other. Either way presents computational challenges.

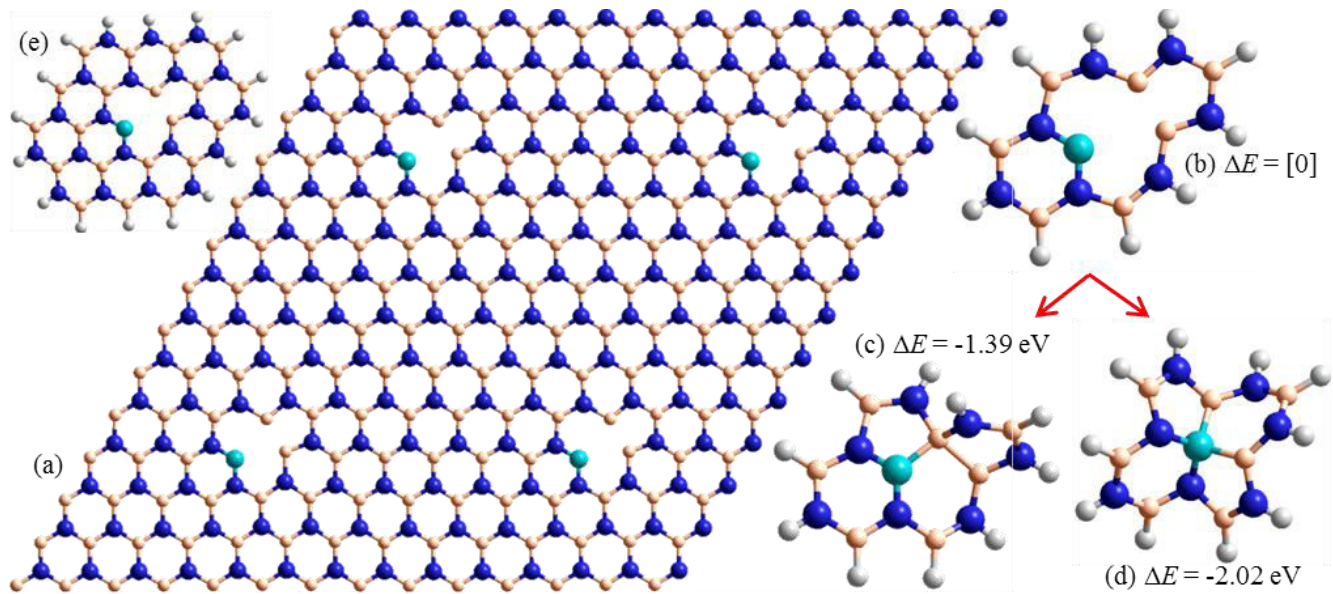

**Figure 1.** The $V_NC_B$ defect in which a nitrogen vacancy is neighbored by a carbon substituting boron is represented as either (a) a lattice of periodic defects or (b) a model compound with $C_{2v}$ symmetry. On relaxation below $C_{2v}$ symmetry, the model compound forms 3-D structures (c) and (d). An expanded model compound is (e). Cyan- carbon, blue- nitrogen, peach- boron, white- hydrogen.

An important feature is that DFT and TDDFT can suffer various severe failings depending on the type of density functional used and the nature of the state being considered,[54-61] demanding that great care be taken. Here we consider the $V_NC_B$ defect in 2-D h-NB nanoflakes that occurs if a nitrogen vacancy has adjacent to it a carbon atom substituting a boron atom. This system is chosen as it shows considerable complexity, with many unresolved issues concerning interpretation of experimental data,[22-24,62-63] that may form a paradigm for the general understanding of the associated phenomena.[7-9,12-14,64-65] DFT and TDDFT results for 25 states of a model compound and for a periodic molecular layer are compared to each other and to MRCI, CASPT2, CCSD(T), CCSD, and EOMCCSD calculations, seeking efficient and reliable methodologies. Features considered include: relative state energies at vertical and adiabatic geometries, free energy corrections to these energies, simple orbital-energy based methods of estimating them, the band widths of photoabsorption and photoemission spectra, and the effect of choice of density-functional type, all discussed in a context of the chemical forces operative at defect sites and the open-shell nature of defect singlet states. Most DFT calculations use the HSE06 density functional[66-67] as this hybrid density functional is the highest-rung functional commonly available in periodic-lattice codes (e.g., this is the case VASP[68-69]). We also briefly consider its related more primitive functional based on the generalized-gradient approximation (GGA), PBE,[70] and in some detail also the advanced long-range corrected functional CAM-B3LYP,[61,71-72] a functional of the type required for modeling spectroscopic problems involving charge transfer.[57,61,71-76] TDDFT calculations need to be performed based on a reference wavefunction and so we also focus on the critical importance of this choice.

## 2. COMPUTATIONAL METHODS

DFT and TDDFT single-point energy calculations, geometry optimizations, normal mode vibrational frequency calculations, and free-energy calculations were performed on a model compound (Fig. 1b) displaying the $V_NC_B$ defect by Gaussian-09[77] using the HSE06 and CAM-B3LYP[71-72] density functional[66-67] with the cc-pVTZ basis set;[78] some single-point energy calculations are also performed using the PBE.[70] Parallel *ab initio* CCSD(T), MRCI, and CASPT2 calculations were also performed using MOLPRO[79] using the same basis set; the internally contracted "MRCIC"[80] and "RS2C"[81] variants were used. EOMCCSD calculations were performed using Gaussian16,[82] with results using closed-shell reference states agreeing to parallel MOLPRO calculations always to better than 0.01 eV.

HSE06 DFT geometry optimizations on this model compound were also performed using VASP,[68-69] as well as calculations on a 2-D material embodying periodic replications of the $V_NC_B$ defect (Fig. 1a). In the VASP calculations, PAW pseudopotentials[83] (Version 5.4; files dated N: "08Apr2002", C: "08Apr2002", B: "06Sep2000", and H "15Jun2001") were used with a plane-wave cuttoff of 350 eV at the Γ-point of large supercells of size 12 × 20 × 20 Å for the model compound and a 10 × 10 supercell of h-BN, with a N-B bond length of 1.452 Å as optimized for the bulk material; a 30 Å interlayer separation is set to ensure minimal interaction.

## 3. RESULTS AND DISCUSSION

### a. Overview of defect nuclear and electronic structure.

The conceived structures[22-23] of the $V_NC_B$ defect in a periodic lattice and in a model compound are depicted in Fig. 1a and Fig. 1b, respectively. These structures have all atoms planar with local $C_{2v}$ symmetry around the defect. Using the standard axis conventions for planar molecules,[84-85] the irreducible representations of this point group are $a_1$ and $b_2$ (of σ type) and $b_1$ and $a_2$ (of π type); note that a variant definition is possible that is sometimes used in which $b_1$ and $b_2$ are interchanged. However, the ground state or any excited state could show distortions from $C_{2v}$ symmetry, with the defect perhaps distorting from planarity, forming *e.g.* the structures shown in Fig. 1c-d. Through use of $C_{2v}$ descriptors, we present a simple understanding of the basic spectroscopic properties of this defect and its propensity to undergo such distortions.

The model compound shown in Fig. 1b has one ring of h-BN atoms surrounding the defect. How this ring is terminated determines the overall electronic structure, and in the figure hydrogen atoms are used at all sites. Substitution of $BH_2$ and/or $NH_2$ for the hydrogens placed on the $C_{2v}$ axis would allow variation between possible quinonoid, semiquinoid, and benzenoid model compounds. Owing to this complexity, only a model compound containing a full second surrounding atomic layer could prove to be large enough to accurately describe the defect, but we use this smaller compound as it is adequate for benchmarking. Neverthelsss, we find later that results for the model compound and the periodic structure are quite similar. Some calculations are also performed on the expanded two-ring model compound shown in Fig. 1e.

The electronic structure of the defect is considered in terms of six defect orbitals interacting with the valence and conduction bands of the h-BN layer.[23] These comprise one σ-type orbital and one π-type orbital taken from the two defect boron atoms and the defect carbon atom. In Fig. 2. are shown the six molecular orbitals with highest defect-orbital character; a feature of immediate note is that these all embody significant delocalization into the surrounding material and so any classification scheme can be only qualitatively descriptive. The σ molecular orbitals are named $1a_1$, $2a_1$, and $b_2$, whilst the π orbitals are named $1b_1$, $2b_1$, and $a_2$. We adopt this notation as it is invariant to the number of atoms used to model the defect, applying to both model compounds and model 2D slabs.

Each set of orbitals takes on a form similar to that found in 3-centre 2-electron bonds, usually described in terms of "bonding" orbitals ($1a_1$ and $1b_1$), "non-bonding" orbitals ($2a_1$ and $2b_1$), and "antibonding" orbitals ($b_2$ and $a_2$), but delocalization of orbitals into the surrounding material significantly distorts their shapes. Further, those labels are misleading as the C-B and B-B distances are large, possibly beyond the singlet-triplet instability point of the "bonds" that could be driven by occupancy of the "bonding" orbitals. Hence the energy differences between the various orbital combinations are small. HSE06 calculations place three of these six molecular orbitals within the band gap of h-BN, with the lowest-energy orbital ($1a_1$) being within the valence band and the two antibonding orbitals within the conduction band. In neutral $V_NC_B$, four electrons are available to occupy the six defect-site orbitals. Mostly two electrons always occupy the deep lying $1a_1$ orbital so that the location of the other two electrons typically defines the nature of the low-energy electronic states. However, to describe qualitatively all low-energy states of $V_NC_B$ of possible current interest, we also include two highly-lying h-BN valence-band orbitals and two conduction-band orbitals in SI Table S1. The greater character that a transition has of such orbitals, the less appropriate are defect cluster models. While the presented situation is quite complex, the issues raised will be apt for general situations involving defects in 2D materials, making $V_NC_B$ a paradigm example.

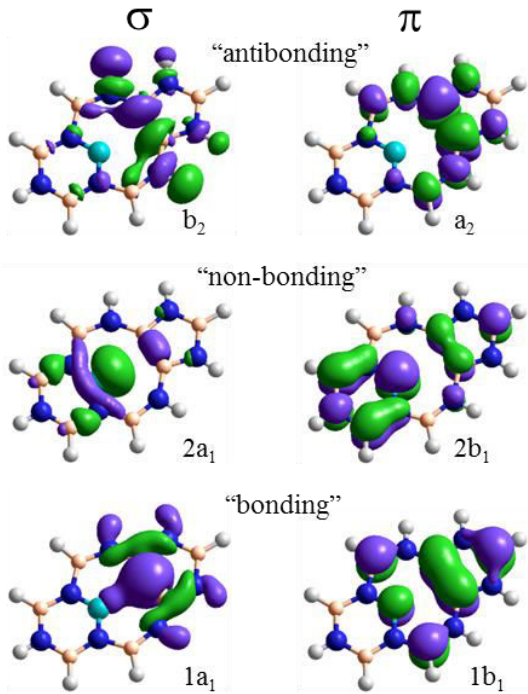

**Figure 2.** Critical orbitals largely associated with the defect site in $V_NC_B$.

### b. Calculations at a single reference geometry. [This section is completely rewritten]

The energies of some 25 low-energy electronic states of the model compound are compared in Table 1, all evaluated at the same test geometry given in Supporting Information (SI). This geometry was obtained by optimizing the structure of the 2-D layer for the $(1)^3B_1$ state, cutting out the model compound and then adding and optimizing terminating hydrogen atoms. HSE06 predicts that $(1)^3B_1$ is the lowest-energy state at this geometry whilst CAM-B3LYP, CCSD(T) and MRCI predict instead that $(1)^1A_1$ is lower in energy. In the table, all energies relate to that of $(1)^1A_1$.

In SI table S1, the dominant orbital occupancies for the 25 states considered are listed. Proper inclusion of the orbitals listed, as well as other significantly coupled ones, is critical to any calculation. In particular, the choice of active space presents a significant practical issue in performing MRCI and related calculations, particularly in situations such as this where many different orbitals contribute to the low-energy states of interest. As detailed in SI Table S1, 5 different active spaces were used in MRCI calculations, with the results in Table 1 being the obtained averages, with the standard deviations of the whole data set being 0.07 eV.

**Table 1. Energies (in eV) of various states of the $V_NC_B$ model compound (Fig. 1b) with respect to $(1)^1A_1$, evaluated at the test geometry.**[a]

| State | MRCI[b] | CASPT2[c] | CCSD | CCSD(T) | EOM-CCSD-S | EOM-CCSD-T | HSE06 VASP | HSE06 G09 | HSE06 TD-S | HSE06 TD-T | CAM | CAM TD-S | CAM TD-T |
|---|---|---|---|---|---|---|---|---|---|---|---|---|---|
| $(2)^1A_1$ |  | 2.30 | 2.73 | 2.03 |  |  | 2.06 | 2.11 |  |  | 2.61 |  |  |
| $(1)^1B_1$ | 1.45 | 0.94 |  |  | 1.25 |  |  | 0.63 | 0.62 |  | 0.89 | 0.91 |  |
| $(2)^1B_1$ |  | 2.98 |  |  | 3.44 |  |  | 2.57 | 2.61 |  | 3.05 | 3.06 |  |
| $(3)^1B_1$ |  | 4.24 |  |  | 4.65 |  |  |  | 3.62 |  |  | 4.28 |  |
| $(4)^1B_1$ |  | 4.68 |  |  |  |  |  |  | 4.96 |  |  | 5.53 |  |
| $(1)^1B_2$ | 5.19 | 4.22 |  |  | 4.79 |  |  | 3.89 | 3.96 |  | 4.36 | 4.56 |  |
| $(2)^1B_2$ | 5.15 | 4.37 |  |  | 4.62 |  |  | 3.79 | 3.84 |  | 4.34 | 4.40 |  |
| $(1)^1A_2$ | 3.28 | 3.03 |  |  | 3.36 |  |  | 2.34 | 2.22 |  | 2.74 | 3.01 |  |
| $(2)^1A_2$ |  |  |  |  | 4.36 |  |  |  | 3.37 |  |  | 4.02 |  |
| $(1)^3A_1$ | 3.96 | 3.50 | 3.97 | 3.66 |  | 4.01 | 3.09 | 3.17 |  | 3.14 | 3.63 |  | 3.62 |
| $(2)^3A_1$ |  |  |  |  |  |  |  |  | 4.17 | 4.25 |  | 4.87 | 4.92 |
| $(3)^3A_1$ |  |  |  |  |  |  |  |  | 4.37 | 4.49 |  | 5.16 | 5.23 |
| $(4)^3A_1$ |  |  |  |  |  |  |  |  | 4.69 | 4.81 |  | 5.19 | 5.45 |
| $(5)^3A_1$ |  | 5.22 | 5.57 | 5.15 |  | 5.07 |  | 4.75 |  | 5.05 | 5.30 |  | 5.34 |
| $(1)^3B_1$ | 0.32 | 0.02 | 0.24 | 0.24 | 0.25 | [0.24] | -0.20 | -0.16 | -0.59 | [-0.16] | 0.09 | -0.61 | [0.09] |
| $(2)^3B_1$ |  | 2.34 |  |  | 2.63 | 3.01 | 1.99 | 2.07 | 1.80 | 2.10 | 2.52 | 2.20 | 2.62 |
| $(3)^3B_1$ |  | 4.33 |  |  |  |  |  |  | 3.49 | 3.57 |  | 4.16 | 4.16 |
| $(4)^3B_1$ |  |  |  |  |  |  |  |  | 4.49 |  |  | 5.00 | 5.04 |
| $(5)^3B_1$ |  | 5.22 |  |  |  |  |  |  | 4.49 |  |  | 5.26 | 5.33 |
| $(1)^3B_2$ |  |  | 4.19 | 4.01 | 4.21 | 4.59 |  | 3.55 | 3.35 | 3.57 | 3.91 | 3.75 | 4.21 |
| $(2)^3B_2$ | 4.96 | 4.66 | 4.74 | 4.42 | 4.60 |  |  | 3.91 | 3.76 | 3.95 | 4.46 | 4.32 | 4.97 |
| $(1)^3A_2$ | 3.27 | 2.42 | 3.01 | 2.85 | 3.32 | 3.18 | 2.23 | 2.29 | 2.15 | 2.26 | 2.70 | 2.94 | 2.83 |
| $(2)^3A_2$ |  |  |  |  | 4.33 | 4.29 |  |  | 3.29 | 3.38 |  | 3.94 | 3.89 |
| $(3)^3A_2$ |  | 3.96 |  | 5.25 | 4.65 |  |  |  | 4.05 |  | 3.96 | 4.78 | 4.76 |

[a] see SI for geometry used. CAM ≡ CAM-B3LYP; EOM-CCSD and TDDFT calculations are preformed from both the $(1)^1A_1$ reference state ("-S") and the $(1)^3B_1$ reference state ("-T"). [b] Average of 5 calculations using different active spaces, ±0.07 eV. [c] Average of 2 calculations using different active spaces, ±0.3 eV.

This value provides a basic estimate of the precision of the method. Similarly, two active spaces are used in the CASPT2 calculations, with the standard deviation increasing to 0.3 eV. That the difference found between the two methods is so large is indicative of strong electron correlation within the defect.

DFT calculations are also subject to implementation variations; to investigate this, results for 6 states are compared in Table 1, obtained using Gaussian-09, applying the cc-PVTZ basis set, and using VASP, applying PAW pseudopotentials and a truncated plane-wave basis set. On average, the cc-pVTZ calculations place the excited states higher in energy by 0.06±0.02 eV. Errors of this order thus represent the basic precision of the DFT calculations and are similar in magnitude to those associated with MRCI active-space choice.

Table 2 provides a comparison of average and standard-deviation differences in excited state energies obtained comparing the various computational results presented in Table 1 to each other. The second column of Table 2 crudely summarizes the results, providing an overall estimate of errors expected from various DFT approaches, standardized against the *ab initio* methods. For some states, these error estimates are within the precisions discussed above for calculation evaluation, whereas in other situations the DFT results differ on average by up to 1 eV. We proceed by considering the properties of the different electronic states involved, as well as those of the defferent DFT and *ab initio* methods used to model them.

*(i) Triplet states.* Calculations of energy differences within the triplet-state manifold are perhaps the ones that are least prone to systematic error, and one would expect the CCSD(T) and DFT results both to be highly reliable. For triplet states, according to the Gunnarsson-Lundqvist (extended Kohn-Sham[86]) theorem,[87] DFT energies can be evaluated directly for the lowest-energy state specified by any occupancy of each spatial symmetry, and the same applies to CCST(T). Sometimes higher-energy states can also be identified, provided that convergence criteria are set course enough.

**Table 2.** Corrections to add to DFT calculated transition energies from $(1)^1A_1$ (second column, in eV), determined from averaged differences in transition energies predicted by the methods listed in the rows compared to those in the later columns, for the $V_NC_B$ model compound (Fig. 1b), evaluated at a test geometry.[a]

| method | Corr. | MRCI | CASPT2 | CCSD | CCSD(T) | EOM-CCSD-S | EOM-CCSD-T | HSE06 closed | HSE06 open | HSE06 trip. |
|---|---|---|---|---|---|---|---|---|---|---|
| CASPT2 | | 8<br>−0.6±0.3 | | | | | | | | |
| CCSD | | 4<br>−0.1±0.1 | 7<br>0.5±0.4 | | | | | | | |
| CCSD(T) | | 4<br>−0.3±0.2 | 7<br>0.1±0.3 | 8<br>−0.3±0.2 | | | | | | |
| EOM-CCSD-S | | 7<br>−0.2±0.2 | 10<br>0.4±0.2 | 4<br>0.1±0.2 | 4<br>0.2±0.2 | | | | | |
| EOM-CCSD-T | | 3<br>−0.0±0.1 | 5<br>0.4±0.3 | 5<br>0.0±0.3 | 5<br>0.2±0.2 | 5<br>0.1±0.2 | | | | |
| HSE06 closed | 0.0 | | 1<br>−0.2 | 1<br>−0.6 | 1<br>0.1 | | | | | |
| HSE06 open | 1.0 | 4<br>−1.1±0.2 | 5<br>−0.5±0.1 | | | 5<br>−0.8±0.1 | | | | |
| HSE06 trip | 0.7 | 4<br>−0.8±0.2 | 7<br>−0.3±0.2 | 7<br>−0.8±0.2 | 7<br>−0.5±0.1 | 5<br>−0.7±0.2 | 6<br>−0.7±0.3 | | | |
| HSE06 TD-S | 0.9 | 7<br>−1.1±0.2 | 13<br>−0.5±0.3 | 4<br>−0.9±0.1 | 4<br>−0.7±0.1 | 13<br>−0.9±0.1 | 6<br>−1.1±0.2 | | 5<br>0.0±0.1 | 5<br>−0.2±0.1 |
| HSE06 TD-T | 0.5 | 4<br>−0.8±0.2 | 8<br>−0.3±0.3 | 7<br>−0.7±0.3 | 7<br>−0.5±0.2 | 6<br>−0.7±0.2 | 8<br>−0.8±0.4 | | | 8<br>0.0±0.1 |
| CAM | 0.4 | 8<br>−0.5±0.2 | 13<br>0.1±0.3 | 8<br>−0.3±0.1 | 8<br>0.1±0.2 | 10<br>−0.3±0.2 | 6<br>−0.3±0.3 | 1<br>0.5 | 5<br>0.4±0.1 | 8<br>0.5±0.1 |
| CAM TD-S | 0.5 | 7<br>−0.6±0.2 | 13<br>0.0±0.4 | 4<br>−0.4±0.3 | 4<br>−0.3±0.4 | 13<br>−0.4±0.2 | 6<br>−0.6±0.2 | | 5<br>0.5±0.1 | 5<br>0.2±0.4 |
| CAM TD-T | 0.0 | 4<br>−0.3±0.2 | 9<br>0.2±0.3 | 7<br>−0.2±0.2 | 7<br>0.1±0.2 | 6<br>−0.1±0.3 | 8<br>−0.3±0.2 | | | 8<br>0.6±0.2 |

| | HSE06 TD-S | HSE09 TD-T | CAM | CAM TD-S |
|---|---|---|---|---|
| HSE06 TD-T | 10<br>0.2±0.1 | | | |
| CAM | 10<br>0.5±0.1 | 8<br>0.4±0.2 | | |
| CAM TD-S | 19<br>0.6±0.2 | 11<br>0.4±0.3 | 10<br>−0.1±0.3 | |
| CAM TD-T | 11<br>0.8±0.2 | 14<br>0.6±0.2 | 8<br>0.1±0.2 | 12<br>0.2±0.3 |

[a] Listed are the number of states used in each comparison, the average difference (in eV), and its standard deviation (in eV). CAM ≡ CAM-B3LYP; HSE06 direct energy calculations for closed-shell singlets, open-shell singlets, and triplet states are shown separately; EOM-CCSD and TDDFT calculations are preformed from both the $(1)^1A_1$ reference state ("-S") and the $(1)^3B_1$ reference state ("-T").

While Tables 1 and 2 presents state energies relative to $(1)^1A_1$, SI Table S2 depicts the higher-energy triplet states relative to $(1)^3B_1$ instead and so as to focus only on properties of the triplet-state manifold. Comparisons are made for direct DFT calculations using the PBE, HSE06, and CAM-B3LYP density functionals. HSE06 performs best with an average difference to CCSD(T) of -0.10±0.07 eV; CAMB3LYP shows its customary over-estimation[60-61] of energies by ca. 0.2 eV, with an average difference of 0.16±0.12 eV, while PBE results are very poor at -0.90±0.29 eV. Clearly PBE is inappropriate to the study of defect sites and we do not investigate this method further. In addition, comparisons are also made to TDDFT results obtained using direct calculations for $(1)^3B_1$ as the reference state, returning similar differences of -0.07±0.20 eV for HSE06 and 0.32±0.21 eV for CAM-B3LYP, with the variability of the results increasing somewhat compared to direct DFT calculations for each individual state.

In addition, Table S2 also compares results within the triplet manifold for MRCI, CASPT2, CCSD, and EOMCCSD to those from CCSD(T). Average differences are of order 0.3 eV, indicating variability in the quality of the treatments of the 6 excited states considered. In principle, the CCSD(T) calculations are expected to perform well as they contain a consistent treatment of triples excitations, but the MRCI and CASPT2 approaches better include any reference-state multi-reference character. In the absence of MRCI calculations including full inclusion of triples excitations, it is difficult to make an estimate of likely errors based on convergence of the *ab initio* results towards the exact answer for the basis set used, but combining the *ab initio* and DFT results is sugges-tive that triples excitations are more important than multi-reference character amongst the triplet manifold. A conservative analysis is that the CCSD, MRCI and EOMCCSD methods can only determine state energies to an accuracy of on average 0.3 eV. EOMCCSD(T) is an additional method that should prove useful in these applications.[34]

*(ii) Closed-shell singlet states.* Only two closed-shell singlet states are considered, $(1)^1A_1$ and $(2)^1A_1$. Of these, $(1)^1A_1$ is of critical importance as most calculations predict it to be the ground state. Typically, closed-shell singlet states are the easiest to calculate as they are often dominated by single-reference character and lead to chemical stability, but this scenario is unlikely to apply to defect states. In $(1)^1A_1$, the $2a_1$ orbital (see Fig. 2 and SI Table S1) is doubly occupied, whereas, in $(2)^1A_1$, $2b_1$ is doubly occupied instead. Both involve electron delocalization over atomic centers that are a long way apart, an unusual property that inherently introduces multi-reference character into wavefunctions. The inter-atomic interactions are barely strong enough to support this delocalization; to allow the electrons to localize on individual atomic centers, the mixing in of double excitations, from amongst in particular the critical defect orbitals shown in Fig. 2, must occur. In general in molecules, this issue becomes critical at bond-broken geometries beyond the singlet-triplet instability point,[56] leading to poor results for DFT ground states and catastrophic failure of TDDFT for excited states. This effect is known as "static electron correlation" and is a distinctly different phenomenon to the dynamical correlation effects addressed in different ways by the DFT and *ab initio* calculations.

Whereas SI Table S2 shows excellent agreement between DFT results and CCSD(T) calculations for energy gaps within the triplet manifold, Tables 1 and 2 report energy gaps with reference to $(1)^1A_1$. For HSE06, the average deviation from CCSD(T) for the triplet-state energies becomes -0.5±0.1 eV (Table 2) compared to -0.10±0.07 (Table S2) for relative energies within the triplet manifold. This is indicative of the serious issues involved in obtaining a realistic electronic structure for $(1)^1A_1$. CAM-B3LYP performs much better, however, the error with respect to $(1)^1A_1$ being just 0.1±0.2 eV (Table 2).

All of the CCSD(T) and the DFT calculations are based on single-reference wavefunctions and so are subject to errors associated with the multi-reference nature of $(1)^1A_1$. Insight into the process is obtained by comparing CCSD(T) results to CCSD in Table 2: CCSD(T) systematically includes much more of the multi-reference character that CCSD, with CCSD transition energies being on average 0.3±0.2 eV higher in energy. Further insight into multi-reference character in the CCSD(T) calculations can be gleaned from descriptors such as "t1", "t2", and "d1"[88-89], and these are listed in SI Table S3 for all states considered. For example, "T1" is small for the lowest triplet states of each symmetry, ~0.06, significant for $(1)^1A_1$, 0.18, and sizeable for $(1)^1A_1$, 0.38.

While CCSD and more so CCSD(T) (asymmetrically) include static electron correlation to some extent, DFT largely misses it. If the effect is too severe then DFT calculations yield open-shell singlet states with lower energies than analogous closed-shell states. We tested for this effect and found it not to occur, meaning that DFT and TDDFT should deliver qualitatively useful results, as we find; the question then becomes one of quantitative accuracy.

An alternative *ab initio* computational method to CCSD(T) giving results independent in quality of the degree of multi-reference character is MRCI and its approximation CASPT2. These method embodies exact treatment of exchange and correlation within selected key orbitals, coupled to either variational or perturbation treatments of the dynamic electron correlation. The MRCI calculations that we perform are truncated at the level of singles and doubles expansion and hence do not include the triples contributions identified as significant for the relative energies of the triplet-state manifold, however. The errors listed in Table 2 for HSE06 compared to MRCI are -0.8±0.2 eV, being also -0.6±0.2 eV for CAM-B3LYP; these are very much larger than the ca. 0.3 eV magnitude errors associated with neglect of triples excitations and emphasize the failure of DFT for the closed-shell $(1)^1A_1$ state.

Indeed, HSE06 predicts that $(3)^1B_1$ should be the ground state at an energy 0.16 eV lower than that of $(1)^1A_1$, owing to its poor treatment of $(1)^1A_1$, whereas MRCI, CCSD, CCSD(T), and EOM-CCSD all predict it to be ca. 0.25 eV in energy higher. For this result, of significant immediate concern regarding interpretation of spectral data for hBN,[24] the *ab initio* methods yield robust results while the DFT calculations suffer from significant systematic errors. CAM-B3LYP more realistically predicts the triplet state is less favored by 0.09 eV (Table 1), whereas PBE predicts it to be more stable by 0.59 eV (SI Table S2); these results all understandable for first-principles understanding of the nature of the density functionals involved.[60-61]

TDDFT calculations based on the $(1)^1A_1$ reference state provide another way of calculating properties of the triplet-state manifold. SI tables S2 highlights these in terms of their ability to predict properties from purely within the triplet-state manifold: HSE06 performs very well, with average errors compared to CCSD(T) of 0.16±0.2 eV, similar to the direct DFT results of -0.10±0.07 eV and the $(1)^3B_1$-reference TDDFT results of -0.07±0.20 eV. However, analogous CAM-B3LYP result fail badly, the error being 0.76±0.18 eV compared to 0.16±0.12 eV (direct) and 0.32±0.21 eV (triplet reference). The results serve to stress the great danger in performing DFDFT calculations based on closed-shell reference functions when the ground state is actually multi-reference in nature.[56] Such approaches should not be applied to model defect-state spectroscopy. There will be a flow-on of this result to related DFT methods for improving ground-state performance such as the random-phase approximation (RPA).

Associated with strong multi-reference ground-state character is often the presence of low-energy doubly excited electronic states. Indeed, $(2)^1A_1$ is predicted by all computational methods to be the second-lowest energy singlet excited state, the lowest-energy singlet excited state being the open-shell state $(1)^1B_1$ for which $(2)^1A_1$ is the associated doubly excited state from $(1)^1A_1$ (see SI Table S1). The difference between the CCSD(T) energy of 2.03 eV and the CCSD energy of 2.73 eV (Table 1) is very large and suggestive of the much better treatment of static electron correlation that CCSD(T) affords. HSE06 and CAM-B3LYP predict 2.11 eV and 2.61 eV, respectively. TDDFT and EOMCCSD inherently misrepresent doubly excited states and are not suitable for their search.

*(iii) Open-shell singlet states.* Many examples of open-shell singlets are reported in Table 1. Calculation for these states are even more complex than those for closed-shell singlets as they are *intrinsically* open-shell in nature, being represented at the simplest level as an equally weighted sum of two Slater determinants. This minimalistic treatment of static electronic correlation is provided by the MRCI, CASPT2, EOM-CCSD, and TDDFT methods but not but by CCSD, CCSD(T), or direct DFT approaches. An *ab initio* extension of CCSD(T) to include multi-reference effects is available,[40-41] as is a successful empirical extension of DFT;[42] also, methods that merge the two approaches that are applicable to periodic materials are becoming available.[43-53] However, such approaches are poorly available and rarely applied in practice, despite their desirability.

In liu of a first-principles approach, direct DFT calculations of open-shell singlet states are usually performed using an empirical multi-reference ansatz: the energy of spin-contaminated states calculated by setting the number of spin-up and spin-down electrons to be the same whilst setting the orbital occupancy (see SI Table A1) of an open-shell excited state is assumed to be the *average* energy of the singlet and triplet states.[90] By also calculating the energy of the triplet state at the same geometry, the energy of the singlet state can thus be estimated. In principle, forces can be determined using the same ansatz and hence geometries optimized, but in practice often only the energy of the spin-contaminated state is optimized. As spin contaminated states violate the Gunnarsson-Lundqvist theorem,[87] they are not valid solutions to DFT and hence any meaning associated to them is purely empirical. The possible consequences of this for the complex states manifest at defect sites remains unknown.

Tables 1 and 2 depict relatively good agreement between MRCI and EOMCCSD calculations for all states, but specifically, for the 4 open-shell singlet states treated by both methods, the average EOMCCSD difference to MRCI is only −0.26±0.27 eV, despite the multi-reference nature of the states. This is small compared to the differences between *ab initio* and DFT results: 1.2±0.2 eV (direct DFT) and 1.1±0.2 eV (TDDFT) for HSE06, with also 0.6±0.2 eV (direct DFT) and 0.7±0.2 eV (TDDFT) for CAM-B3LYP. For both density functionals, the use empirical direct DFT scheme yields results very similar to TDDFT (0.0±0.1 eV for HSE06 and 0.1±0.3 eV for CAM-B3LYP from Table 2), so this approach is not the cause of the problem. Rather, the problems are intrinsic ones associated with the poor treatment of multi-reference character in traditional DFT.

**b. Optimized Geometries.**

Some other issues concerning calculation of the properties of defect states are presented in Table 3. In this table, all energies are reported at geometries individually optimized using HSE06 (see SI), with most columns referring to the model compound whilst the last two column are for the 2-D material. CAM-B3LYP, CCSD(T) and EOMCCSD single-point energies are also evaluated at these structures. Geometry optimization is seen to enhance the effects found previously at the reference geometry. Direct DFT calculations, TDDFT, and CCSD(T) tell a consistent story for the triplet manifold, whilst empirical DFT and TDDFT give similar results for the open-shell singlet states that are far removed from the EOMCCSD ones. As the ground state of a defect is determined by the adiabatic differences in state energies evaluated at fully optimized geometries, reliable DFT predictions of defect properties becomes can become difficult and human-intuition dependent. Understanding how the errors in the DFT calculations change with geometry requires understanding the operational chemical forces and their consequences concerning the shapes of the potential-energy surface manifolds.

**c. Chemical forces acting to heal the defect.**

Normal-mode analyses were performed at all HSE06 excited-state structures optimized by TDDFT, with the nature of the optimized stationary points determined. Not all optimized structures are found to be local minima on the potential-energy surface, and the symmetry of modes found to break $C_{2v}$ symmetry are listed in Table 3. These modes are mostly out of plane modes that lead to a variety of low-symmetry structures, with most acting to try to rejoin bonds to heal the defect. Most significantly, the lowest-energy state found at $C_{2v}$ symmetry, $(1)^1A_1$, undergoes a displacement in a $b_1$ model leading to a structure of $C_s$ symmetry with a HSE06 energy change of $\Delta E$ = -0.27 eV (see SI), followed by a second displacement to a $C_1$ symmetry (Fig. 1c) at $\Delta E$ = -1.39 eV. However, relaxation of $(2)^1A_1$ leads instead to a lower-energy structure at $\Delta E$ = -2.02 eV (Fig. 1d) that is the lowest-energy structure found by HSE06 for the model compound. In these structures, the molecular plane buckles to bring defect atoms close enough together to form bonds.

The chemical forces driving these transformations will also occur for $V_NC_B$ defects in 2-D and 3-D materials but will be opposed by constraining forces coming from the h-BN layer and the surrounding environment. In considering defect sites, in general strong chemical forces of this nature will arise and the strength by which they are opposed will control the structural and spectroscopic properties of the defect. Optimization of 2-D layers containing the $V_NC_B$ defect results in various states all of which display $C_{2v}$ symmetry, indicating that in this case the restoring forces overpower the bond formation forces to lock in the defect. Under such circumstances, calculations on model compounds are best performed using the full symmetry of the 2-D material, as is done herein.

Some test calculations performed using the expanded 2-ring compound shown in Fig. 1e indicate that the $(2)^1A_1$ state becomes stable to both $a_2$ and $b_1$ distortions compared to the one-ring model, but even in the expanded system $(1)^1A_1$ remains unstable to $b_1$ distortion. How large a model compound needs to be thus remains a key question. In practical calculations on 2-ring models, it may be advantageous to freeze the outer ring atoms at their geometry extracted from a periodic model.

Another feature of note is that read 2D monolayers are subject to buckling that is not represented in small-scale periodic models (e.g., Fig. 1a). How such buckling interacts with the restorative chemical forces will always be a question of interest concerning defect modeling.

Nevertheless, a consequence of the action of restoring forces is that the wide variation in the quality of DFT results as a function of geometry reported for the one-ring model compound in Table 3 will be reduced somewhat. As the reference structure used in the calculations reported in Tables 1 and 2 is typical of that found in other defects,[22] we anticipate that the perceived errors in DFT found at this structure will be typical of what could be expected in general applications to defect spectroscopy.

Table 3. Energies (in eV) of various states of the $V_NC_B$ model compound and related periodic 2-D material (Fig. 1), evaluated at their adiabatic minimum-energy geometries (see SI), as well as the HSE06 free-energy correction $\Delta G_{corr}$, in eV at 298 K, the nature of any imaginary frequencies, the reorganization energies $\lambda$ depicting the width, in eV, of spin-allowed photo-absorption (and approximately PL) spectra from either the $(1)^1A_1$ (singlet) ground state or the $(1)^3B_1$ (triplet) ground state.

| State | Model compound | | | | | | | | | | | 2-D | |
|---|---|---|---|---|---|---|---|---|---|---|---|---|---|
| | HSE06 G09 | HSE06 VASP | HSE06 TDDFT[a] | HSE06 Est.[b] | CCSD(T)[c] | EOM-CCSD[c] | $\Delta G_{corr}$ | Imag. Freq. | $\lambda$ HSE06[ad] | $\lambda$ CCSD[ce] | $\lambda$ CAM-B3LYP[a] | HSE06 VASP | $\lambda$ HSE06 |
| $(1)^1A_1$ | [0] | [0] | [0] | -1.2 | [0] | [0] | [0] | $b_1$[g] | [0] | [0] | [0] | [0] | [0] |
| $(2)^1A_1$ | 1.93 | 2.50 | | 4.28 | 2.21 | | -0.04 | $a_2,b_1$[h] | 3.32 | 2.46 | 2.41 | 2.46 | 1.15 |
| $(1)^1B_1$ | | 1.59 | 1.54 | [1.54] | | 2.30 | 0.00 | none | 0.50 | 0.43 | 0.47 | 1.08 | |
| $(2)^1B_1$ | | | 3.00 | 2.82 | | 3.97 | 0.02 | $b_1$ | 0.56 | 0.34 | 0.35 | | |
| $(1)^1B_2$ | | | 4.55 | 4.36 | | 5.91 | -0.20 | $b_2$ | 0.76 | 0.14 | 0.25 | | |
| $(2)^1B_2$ | | | 3.82 | 3.53 | | 4.40 | -0.11 | $b_1$ | 0.41 | 0.35 | 0.31 | | |
| $(1)^1A_2$ | | | 2.41 | 2.47 | | 3.35 | -0.05 | none | 0.72 | 0.84 | 0.81 | | |
| $(2)^1A_2$ | | | 3.83 | 4.01 | | 4.82 | -0.17 | $b_1$ | 0.77 | 0.80 | 0.80 | | |
| $(1)^3A_1$ | 4.01 | 4.27 | 4.00 | 4.79 | 4.61 | | -0.11 | $a_2,b_1,b_2$ | 0.68 | 0.75 | 0.69 | 3.00 | |
| $(1)^3B_1$ | 0.77[f] | 0.71 | [0.77] | [0.77] | | | -0.04 | none | [0] | [0] | [0] | 0.34 | [0] |
| $(2)^3B_1$ | | 2.47 | 2.70 | 2.05 | 1.21 | | -0.18 | $b_1,b_2$ | 0.19 | | 0.30 | 1.92 | 0.18 |
| $(1)^3B_2$ | 3.81 | | 4.02 | 3.59 | 4.18 | | -0.02 | $a_2,b_1$ | 1.03 | 1.65 | 1.40 | | |
| $(2)^3B_2$ | 3.70 | | 4.24 | 2.76 | 4.04 | | -0.16 | $b_1$ | 0.63 | 0.81 | 0.39 | | |
| $(1)^3A_2$ | 2.38 | | 2.28 | 1.7 | 2.91 | | -0.18 | none | 0.59 | 0.56 | 0.59 | | |
| $(2)^3A_2$ | | | 3.11 | 3.24 | | | -0.25 | $a_2,b_1,b_2$ | 0.41 | | 0.48 | | |
| $(3)^3A_2$ | 4.48 | | 4.47 | 5.5 | | | -0.13 | $b_1$ | 0.46 | 0.54 | 0.38 | | |

[a] singlet states from $(1)^1A_1$ reference state, triplet states from $(1)^3B_1$ reference state. [b] Crude estimate based on DFT orbital energy differences at the $(1)^1A_1$ geometry added to the TDDFT energies of $(1)^1B_1$ and $(1)^3B_1$. [c] evaluated at HSE06 geometries optimized using DFT or TDDFT. [d] TDDFT except direct DFT for $(2)^1A_1$. [e] EOMCCSD for open-shell singlets else CCSD(T). [f] 0.72 eV using 2-ring structure Fig. 1e. [g] also imaginary for the 2-ring compound in Fig. 1e. [h] local minimum for 2-ring compound in Fig. 1e.

**d. Widths of photoabsorption and photoluminescence spectra.**

How the chemical forces act to heal the defect in its various electronic states is revealed through measurements of photoabsorption and photoluminescence spectra. To model such spectra, calculations need to properly represent both these forces and the restoring ones operating in 2-D and 3-D structures.

While high-resolution spectra provide detailed information concerning the motion of all atoms, the width of the spectral band alone provides the most important qualitative information, specifying the reorganization energy associated with the transition. Large atomic motions in high-frequency modes results in broad bands; alternatively, if the bonding is insensitive to nuclear motion, then spectra will be sharp. Broad spectra are also produced if orbitals from the valence or conduction bands of the h-BN are involved as interactions within these bands create a multitude of electronic transitions out of a single one. Sharp transitions are typically easier to observe and hence dominate experimental studies. Amongst defects, there is a search for those that have defect orbitals lying within the bandgap that have allowed spectroscopic transitions.[22] Photoluminescence usually becomes the focus as it easier to observe than photoabsorption as defects are usually present only in low concentration, with an added bonus often arising if the emitting state is different to the absorbing state and has different symmetry such that the absorption and emission polarizations vary.[91]

Table 3 compares results from TDDFT (HSE06 and CAM-B3LYP functionals) with those form EOMCCSD (singlet states) and CCSD(T) (triplet states) for the bandwidths of photoabsorption spectra originating from either $(1)^1A_1$ or $(1)^3B_1$. These widths are given as the reorganization energies $\lambda$ calculated as the energy change of the excited state in going from its adiabatic minimum-energy geometry to that when vertically excited from the ground state. This corresponds to the second moment of observed absorption spectra. In the harmonic approximation to the ground-state and excited-state potential energy surfaces, if the form of the normal modes does not change then the bandwidth for photoemission corresponds to that for photoabsorption, a commonly good approximation in molecular spectroscopy but one that can also fail badly, e.g., for chlorophylls.[92-93] In practice, reorganization energies become susceptible to vibronic coupling effects that can dramatically change the shapes of excited-state potential-energy surfaces. This occurs as the influence of vibronic coupling depends strongly on the energy gaps between excited states, demanding global accuracy across the excited state manifold if computational methods are to satisfactorily predict absorption and emission bandwidths.[94]

The results presented in Table 3 show wide variations between reorganization energies calculated using HSE06 and either EOMCCSD/CCSD(T) or CAM-B3LYP. Small numbers occur whenever the singlet/triplet excited state has a similar geometry to the singlet/triplet ground state. For $(2)^1B_2$, the difference is stark: 0.76 eV for HSE06 vs. 0.25 eV for CAM-B3LYP and 0.14 eV for EOMCCSD, indicating that the vibronic environments of this state is being perceived very differently by HSE06. For $(1)^3B_2$, HSE06 deviates by the same magnitude but opposite sign, with the largest deviation for HSE06 being 0.9 eV for $(2)^1A_1$. For HSE06, averaging 12 states gives an error of 0.06±0.38 eV whereas for CAM-B3LYP this is only -0.07±0.14 eV.

Most significantly, agreement between all 3 methods is within 0.1 eV is found for the low-energy states $(1)^1B_1$, $(2)^3B_1$, and $(1)^3A_2$ of greatest relevance to the interpretation of observed PL spectra.[22,24] Further, the neglect of restoring forces in these calculations on a model cluster means that the calculated reorganization energies and their sensitivities to the global details of the electronic structure will be overestimated. To explore these effects, HSE06 reorganization energies calculated for the 2-D material are also shown in Table 3. For $(2)^3B_1$, the model compound and the 2-D material have reorganization energies of 0.19 eV and 0.18 eV, respectively, indicating the similarity of the chemical properties of these two model systems.

Insight into the difficulties in modelling the widths of higher-energy transitions comes from noting that, for $(2)^1A_1$, the 2-D material has a large HSE06 reorganization energy of 1.15 eV, indicating that strong chemical forces come into play, but in the model compound this increases to 3.32 eV! Understanding the shapes of excited-state potential-energy surfaces comes down to understanding the conical intersections that these surfaces have with other surfaces. The nature and location of these conical intersections are controlled by subtle differences in energy gaps and state orderings.[94-95] A serious problem with GGA density functionals like PBE and hybrid functionals like HSE06 is that they predict an incorrect asymptotic potential.[96] The error in the asymptotic potential is state dependent and adds to the transition energy, with errors becoming large when significant charge-transfer occurs during an electronic transition. In general, excited-state charge-transfer band energies are predicted to be much too low in energy compared to the charge-neutral states, with this effect having severe consequences as the transition energy increases. Consequences include the prediction by GGA and hybrid functionals that polyacetylene has a triplet ground state,[59] as well as their inability to describe the spectra of porphyrins, chlorophylls, and other systems allowing for extensive charge transfer.[59] Movement of charge to/from defect sites can indeed induce significant long-range charge transfer and so defect sites are susceptible to it. Functions like CAM-B3LYP correct for the errors in the asymptotic potential and hence do not suffer from the chaos that sets in at (nominally) high energies when GGA or hybrid functionals are used, allowing spectral properties of charge-transfer systems to be determined using DFT.[60-61]

**e. Determining the excited states likely to be of relevance.**

Calculation methods like TDDFT and EOMCCSD deliver all desired singly excited states up to some energy at a particular geometry. Based on these results, most states likely to be of interest in some application can usually be determined, the exception being possibly critical doubly excited states which are not treated properly by these techniques. However, direct or empirical DFT calculations, which in practical situations may be all that is feasible, as well as CCSD(T) calculations, require the individual specification of electronic state occupancies as input. Human intuition then can become a controlling element in the identification of the key properties of a defect of interest. This also applies to the search for relevant doubly excited states.

To aid this intuition, it is possible to estimate realistically the energies of many singlet and triplet states given the energy of one excited state determined for each type and the energies of the involved molecular orbitals. All that one needs to do is add orbital energy differences coming from the dominant configurations (SI Table S1) to the known transition energies. Results for this estimation procedure applied at the $(1)^1A_1$ optimized geometry are given in Table 3, where they are compared to the adiabatic energies of the fully optimized electronic states. The estimates are close for the singlet states but range from an underestimation of 1.5 eV to an overestimation of 1.0 eV for the triplet states. Hence to be confident that all low-energy states are properly found by a manual search over a few possibilities, all states at an estimated energy of up to 2 eV above that required should be considered. A simpler estimation procedure using only the orbital energies should be able to predict state order to that type of accuracy as well.[61]

**f. Unexpected differences between DFT outcomes.**

A feature of note is that the empirical DFT calculations for a $^1B_1$ state using VASP converged on an apparent local minimum not reported in the tables. Similar calculations using Gaussian-09 could not be successfully completed as a lower-energy state with the same symmetry occupancy exists to which this program converged instead, despite efforts to converge on the higher-energy state. In the VASP calculations, the occupancy was set for $(2)^1B_1$ (see SI Table S1), but orbital[97] projection of the final results onto the orbitals of $(1)^1A_1$ indicated that the final state was $(1)^1B_1$ instead. How such a local minimum could coexist with the reported local minimum for $(1)^1B_1$ is not clear. However, hassle-free optimization of this state using TDDFT produced a barrierless collapse to the structure of $(1)^1B_1$, suggesting that the result from the empirical DFT calculation was fictitious. This problem with the $^1B_1$ manifold, warns of the dangers of artificially setting non-Aufbau occupancies (in VASP using options "FERWE" and "FERDO") for examining excited states beyond the lowest-energy state of each symmetry occupancy.

**g. Thermal corrections and the Gibbs free energy.**

The purpose of many calculations on defect structures is to predict the state of lowest energy. However, what is observed is the state of lowest *free* energy and so results like those discussed previously need to be corrected for zero-point motion, enthalpy if appropriate, and entropy. Shown in Table 3 are the corrections that need be applied to obtain the Gibbs free energy at 298 K from the otherwise reported total HSE06 electronic energies. These were evaluated from the normal mode analyses previously described, obtained at the TDDFT level. These corrections range over 0.3 eV and are small compared to many other errors and choice variations identified for the $V_NC_B$ defect. In some circumstances, these corrections could be large enough to control the nature of observed defects, however, and should always be included in accurate calculations. It is unlikely that methods variations will significantly influence this contribution.

## 4. CONCLUSIONS

The use of DFT computational schemes to examine the geometrical and spectroscopic properties of defect sites in materials is widespread, with many successes attained. Here, however, we see that for the $V_NC_B$ site in h-BN, comparison of DFT-based results to accurate *ab initio* ones for a model compound indicate that the PBE density function is unsuitable, that for certain properties HSE06 predict very accurate answers but for others may involve large systematic errors of up to 1 eV in magnitude, and that results from CAM-B3LYP suffer less seemingly random anomalies and are generally much more reliable. While only methods containing correction for the asymptotic potential error should be applied to defect sites owing to their likelihood of involving charge-transfer spectroscopy, practical software limitations may limit current applications in periodic systems to uncorrected functionals like HSE06. The implementation of range-corrected hybrid functionals like CAM-B3LYP into solid-state DFT codes is a priority as defect-state spectroscopy is currently a very important research field.

Generally, when using DFT, serious problems occur when states involve significant multi-reference character, especially when comparing the energies of states which differ markedly in this regard. Errors are most pronounced for closed-shell singlet states with "bonding"-type wavefunctions spanning broken bonds. If such a state is used to describe the excited states in a TDDFT calculation, then these errors in the ground state transfer to the excited states produced. At our test geometry, the HSE06 energies of triplet states too predicted to be too low by ca 0.7 eV compared to the energies of closed-shell states, while the energies of open-shell singlet states are underestimates by ca. 1 eV (Table 2). For CAM-B3LYP, the errors are less, 0.3 eV for triplets and 0.5 eV for open-shell singlets, with an average error of 0.4 eV (Table2). The use of TDDFT to calculate triplet state energies based on a triplet-state reference function works well, but closed-shell singlet states with significant multi-reference character cannot be used reliably.

The errors found appear to be minimal when the defect singlet states contain minimal covalent character. Geometrical relaxation tries to heal defects by forming covalent bonds, leading to possible significant enhancement of the error magnitudes described above, with calculated bandwidths for absorption and emission spectra being subject to the same effect. However, in 2-D and 3-D materials, strong restoring forces oppose these chemical forces, minimizing errors anticipated in DFT calculations towards the more basic results observed at our test geometry.

Errors of order 0.3 eV in calculated state energies can be introduced through neglect of zero-point energy and entropy contributions at 298 K. While these are small compared to the other errors in DFT calculations considered, these effects will need to be included in accurate calculation.

Crude estimates of adiabatic state energies obtained using orbital energies at one geometry can significantly underestimate or overestimate the actual energies of states. When selecting states for manual evaluation, one needs to include all states of energy up to 2 eV above that thought to be relevant.

At present, it would appear that the only commonly available reliable approaches for calculating defect states in 2-D and 3-D materials are *ab initio* methods such as CCSD(T), EOMCCSD, CASPT2 and MRCI applied to model compounds chosen large enough so that they adequately represent the extended system in which the defect is found. The recent development[35-40] of efficiently scaling techniques of this type gives potential for expanded applications. However, these methods are still not as accurate as is desired, with variations of on average 0.3 eV being found between different approaches. Improved approaches treating well both multi-reference character and triples excitations are therefore required. On the other hand, the DFT functional found most reliable, CAM-B3LYP, is still not robust enough for general use. However, HSE06 calculations should be adequate for evaluating free-energy contributions and for optimizing geometries. If used without external calibration by *ab initio* methods, DFT results are expected to be very accurate when only triplet states of defects are involved, with the next best scenarios being when either only closed-shell singlet states or else only open-shell singlet states are involved.

A related issue for studies of model compounds or periodic 2-D samples is the need to use systems large enough to minimize errors. Based on our results and the identified strong competition between chemical and restorative forces in controlling defect structure, model compounds intended for accurate modelling should include one or more rings of atoms outside the ring that minimally describes the defect. Use of model compounds that are too small will introduce artefacts owing to the propensity of defect sites to close on themselves to optimize covalent bonding. Also, while typically transitions of interest are defect-localized, this may not always be the case and the appropriateness of cluster models to transitions involving conduction or valence band transitions will always be questionable.

## ACKNOWLEDGMENT

This work was supported by resources provided by the National Computational Infrastructure (NCI), and Pawsey Supercomputing Centre with funding from the Australian Government and the Government of Western Australia, as well as Chinese NSF Grant #1167040630. SA acknowledges receipt of an Australian Postgraduate Award funded by ARC DP 150103317. Funding is also acknowledged from ARC DP 160101301

## ASSOCIATED CONTENT

Provided in supporting information are the coordinates for the molecules and 2-D layers reported.

## AUTHOR INFORMATION

### Corresponding Author

* Mike.Ford@uts.edu.au, Jeffrey.Reimers@uts.edu.au

SYNOPSIS TOC

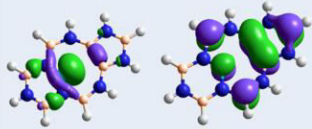